\documentclass{PoS}
\usepackage{amsmath}

\title{Pseudoscalar decays into lepton pairs from rational approximants}

\ShortTitle{Pseudoscalar decays into lepton pairs from rational approximants}

\author{\speaker{Pablo Sanchez-Puertas} \footnote{eprint:~MITP/15-084}
         \thanks{We would like to thank the organizers for their hospitality and the nice atmosphere during the conference.}\\
        PRISMA Cluster of Excellence, Institut f\"ur Kernphysik, Johannes Gutenberg-Universit\"at, Mainz D-55099, Germany\\
        E-mail: \email{sanchezp@kph.uni-mainz.de}}

\author{Pere Masjuan\\
        PRISMA Cluster of Excellence, Institut f\"ur Kernphysik, Johannes Gutenberg-Universit\"at, Mainz D-55099, Germany\\
        E-mail: \email{masjuan@kph.uni-mainz.de}}

\abstract{
The pseudoscalar decays into lepton pairs $P\rightarrow\overline{\ell}\ell$ are analyzed with the machinery of Canterbury approximants, an extension of Pad\'e 
approximants to bivariate functions. This framework provides an ideal model-independent approach to implement all our knowledge of the pseudoscalar transition form 
factors driving these decays, can be used for data analysis, and allows to include experimental data and theoretical constraints in an easy way, and determine a systematic error. 
We find that previous theoretical estimates for these branching ratios have underestimated their theoretical uncertainties. 
From our updated results, the existing experimental discrepancies for $\pi^0\rightarrow e^+e^-$ and $\eta\rightarrow \mu^+\mu^-$ channels cannot be explained unless 
the doubly-virtual transition form factors behavior ---not yet measured--- 
are out of theoretical expectations, which is an interesting result both for anomalous magnetic moments of leptons, and for physics beyond the standard model.
}

\FullConference{The 8th  International Workshop on Chiral Dynamics \\
		 29 June 2015 - 03 July 2015\\
		 Pisa, Italy}

\begin{document}

\section{The process and the puzzle}

\begin{figure}
\centering
   \includegraphics[width=0.5\textwidth]{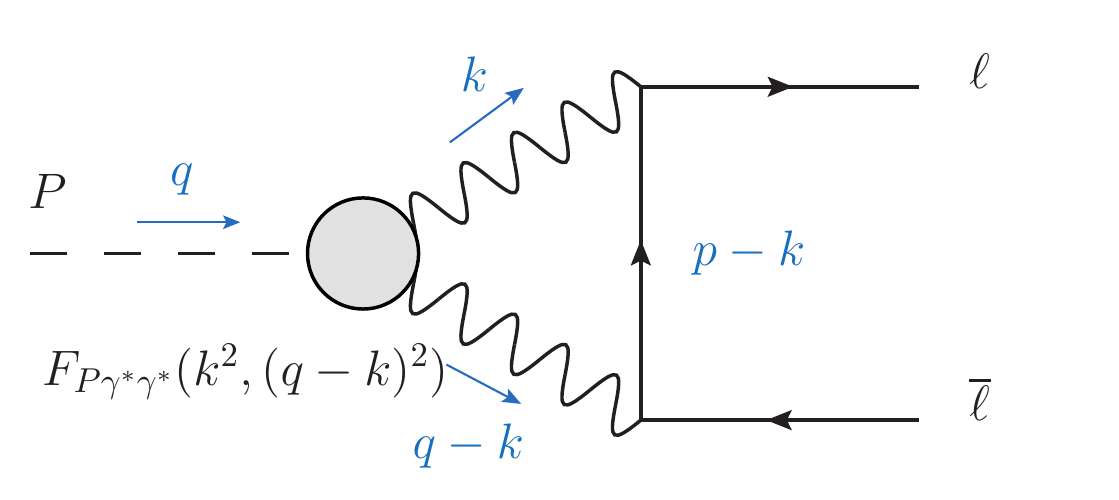}
   \caption{Leading order QED contribution driving $P\rightarrow\overline{\ell}\ell$ decays.}
   \label{fig:FD}
\end{figure}

Pseudoscalar decays into lepton pairs are an interesting playground to test our knowledge of the lowest-lying pseudoscalar mesons. 
The reason being that such processes occurs at the loop level in the standard model (SM) of particle physics (neglecting the tiny $Z^0$-boson contribution). 
Such loop is mediated through an intermediate two-photon state, see Fig.~\ref{fig:FD}, which then requires an accurate knowledge of $P\gamma^*\gamma^*$ interactions 
at all energies, where $P=\pi^0,\eta,\eta'$. This interaction can be parametrized as 
\begin{equation}
i\mathcal{M}_{P\gamma^*\gamma^*} =  i \epsilon_{\mu\nu\rho\sigma} q_1^{\mu}\epsilon_1^{\nu} q_2^{\rho}\epsilon_2^{\sigma}F_{P\gamma^*\gamma^*}(q_1^2,q_2^2),
\end{equation}
where $F_{P\gamma^*\gamma^*}(q_1^2,q_2^2)$ is the double virtual pseudoscalar transition form factor (TFF), which due to Bose symmetry is a symmetric 
function of the photons virtualitites. This function encodes the underlying structure of the pseudoscalar meson,  providing thus a valuable information.
Remarkably, all the measured decays so far~\cite{Abegg:1994wx,Abouzaid:2006kk,Agashe:2014kda}
%
exhibit deviations from the available theoretical predictions, for more details see the preliminary results of this work in Tab.~\ref{tab:res}. 
Since in the SM these branching ratios (BRs) are suppressed by the loop and helicity flip, if the discrepancies persist, they may suggest some new physics scenario, specially 
those of axial or pseudoscalar nature, such as~\cite{Kahn:2007ru,Chang:2008np,Davoudiasl:2012ag}. 
The recent advancements in the field, either from a recent re-evaluation of radiative corrections~\cite{Vasko:2011pi,Husek:2014tna} or TFF 
modelization~\cite{Knecht:1999gb,Dorokhov:2007bd,Dorokhov:2009xs} cannot explain the discrepancies at all. There is however, two aspects  
which have not been fully exploited so far, the treatment of double virtuality and systematic errors. In the present work, we employ the machinery of Pad\'e 
theory extended to the doubly-virtual case, without modeling or prejudice, to account for the TFF description. Our framework provides a powerful tool which can be used 
to analyze experimental data, which would have  then the last word.
The method allows as wekk, for the first time, for a rigorous evaluation of systematic errors, an ingredient never discussed so far.
%

\section{Main properties: low versus high energies}

\begin{figure}
\centering
 \includegraphics[width=0.45\textwidth]{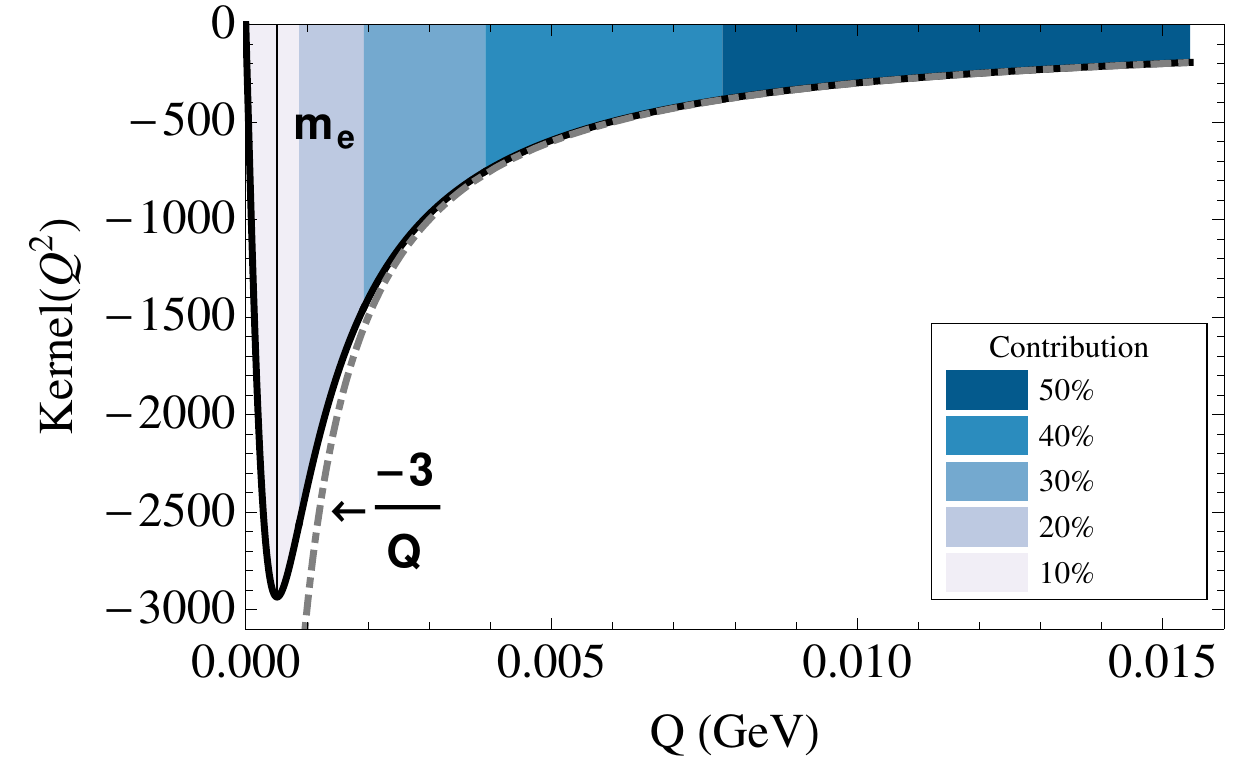}
 \includegraphics[width=0.45\textwidth]{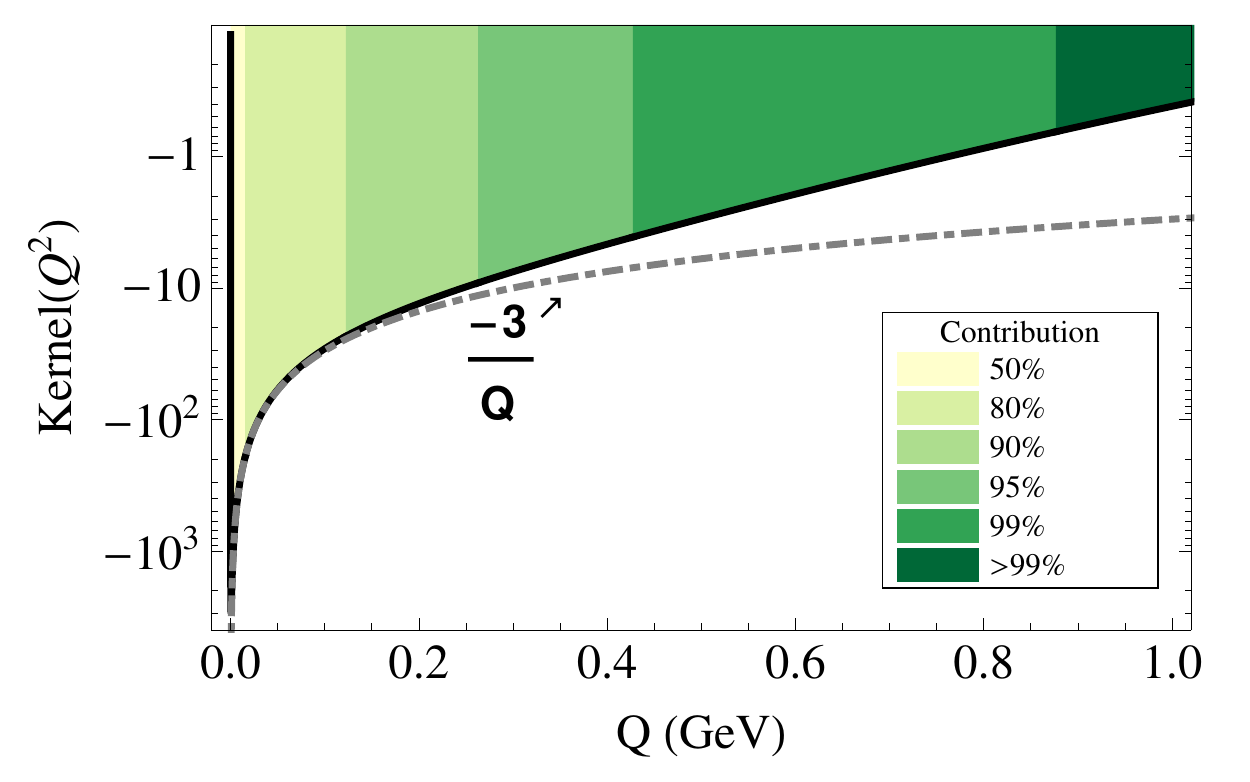}
 \caption{Integral kernel in Eq.~\protect\eqref{eq:loopapp}. The different bands give the relative contribution to the integral.}
 \label{fig:kernel}
\end{figure}

The leading order QED contribution driving $P\rightarrow\overline{\ell}\ell$ decays is depicted in Fig.~\ref{fig:FD} and its amplitude can be expressed in terms 
of the $P\rightarrow\gamma\gamma$ decay as
\begin{equation}
\frac{\textrm{BR}(P\rightarrow\overline{\ell}\ell)}{\textrm{BR}(P\rightarrow\gamma\gamma)} = 2 \left( \frac{\alpha m_{\ell}}{\pi m_P}  \right)^2 \beta_{\ell} |\mathcal{A}(m_P^2)|^2,
\end{equation}
where $\beta_{\ell}(m_P^2) = (1-4m_{\ell}^2/m_P^2)^{1/2}$ is the outgoing lepton velocity, $m_P$ and $m_{\ell}$ are the pseudoscalar and lepton masses, respectively,
and $\mathcal{A}(m_P^2)$ is the loop amplitude 
\begin{equation}
\label{eq:loopint}
\mathcal{A}(q^2) =  \frac{2i}{\pi^2q^2} \int d^4k \frac{ q^2k^2 - (q\cdot k)^2}{ k^2(q-k)^2((p-k)^2-m_{\ell}^2)}\tilde{F}_{P\gamma^*\gamma^*}(k^2,(q-k)^2), 
\end{equation}
that depends on the normalized TFF ($\tilde{F}_{P\gamma^*\gamma^*}(0,0)=1 $) which must be provided at this point. Such object is of paramount importance as otherwise 
the integral would logarithmically diverge.
For the case of light pseudoscalars, i.e., the $\pi^0$, it is still possible to derive further properties from~\eqref{eq:loopint}. The smallness of the $\pi^0$ mass does not 
allow to have any intermediate hadronic states contributing to the imaginary part, which is thereby fully given by the intermediate $\gamma\gamma$ state, 
\begin{equation}
Im(\mathcal{A}(q^2)) = Im(\mathcal{A}_{\gamma\gamma}(q^2)) = \frac{\pi}{2\beta_{\ell}(q^2)}\ln\left( \frac{1-\beta_{\ell}(q^2)}{1+\beta_{\ell}(q^2)} \right).
\end{equation}
This provides a model-independent lower limit $|\mathcal{A}(q^2)| \geq Im(\mathcal{A}_{\gamma\gamma}(q^2))$ known as unitary bound~\cite{Drell}, which for the $\pi^0\rightarrow e^+e^-$ 
decay yields $\textrm{BR}(\pi^0\rightarrow e^+e^-) \geq 4.7\times10^{-8}$. By contrast, the heavier $\eta$ and $\eta'$ mesons admit intermediate hadronic states contributing to the imaginary part. 
For this reason, such bound does not exist for these cases and its use as a reference number is misleading. The numbers provided in Tab.~\ref{tab:res} however can do this 
job, specially for the experimental programs to decide on the hours of data taking. 
In addition, for light pseudoscalars, the heavy scale introduced by the TFF, $\Lambda_{\textrm{TFF}} \gg m_{P} > m_{\ell}$ in Eq.~\eqref{eq:loopint}, allows to approximate 
the loop-integral~\eqref{eq:loopint} as~\cite{Masjuan:2015lca}
\begin{equation}
\label{eq:loopapp}
\mathcal{A}(q^2) = 
\frac{i\pi}{2\beta_{\ell}}L +\frac{1}{\beta_{\ell}}\left( \frac{1}{4}L^2+ \frac{\pi^2}{12} + Li_2\left( \frac{1-\beta_{\ell}}{1+\beta_{\ell}} \right) \right) 
-\frac{5}{4} + \int_0^{\infty}dQ \frac{3}{Q}\left( \frac{m_{\ell}^2}{m_{\ell}^2+Q^2} - \tilde{F}_{P\gamma^*\gamma^*}(Q^2,Q^2) \right),
\end{equation}
where $L= \ln\left[ (1-\beta_{\ell}(q^2))/(1+\beta_{\ell}(q^2)) \right]$, $Li_2$ is the dilogarithm function and $Q$ is the space-like photon virtuality. This kind of approximation, 
in this or any other form, has been widely used in the literature~\cite{Knecht:1999gb,Dorokhov:2007bd,Bergstrom:1983ay}, and is particularly useful for connecting to  
chiral perturbation theory ($\chi$PT), see Ref.~\cite{Knecht:1999gb}. 
The remaining dependence on the TFF is the integral in Eq.~\eqref{eq:loopapp}, where three important properties arise.
First, the relevant TFF region is space-like. Second, the TFF is evaluated at $Q_1^2=Q_2^2$ photon virtualities, making it very sensible to the unknown
doubly-virtual behavior. Third, due to the smallnes of $m_{\ell},m_P$, and the photons propagators, the integrand is extremely peaked at very low energies, saturating 
below $1$~GeV, which requires an extremely precise description for the TFF at these energies, see Fig.~\ref{fig:kernel}, where the function 
$\textrm{Kernel}(Q^2)=3Q^{-1}[m_{\ell}^2/(m_{\ell}^2+Q^2)-\tilde{F}_{P\gamma^*\gamma^*}(Q^2,Q^2)]$ refers to the integral in Eq.~\eqref{eq:loopapp}.  
\\
Given the relevant (low-)energy range, one could try to use $\chi$PT to fully calculate the process. Then, at leading order, the imaginary part of 
diagram in Fig.~\ref{fig:FD} is finite, but its real part diverges and requires a local counterterm $\chi$~\cite{Knecht:1999gb,Savage:1992ac}, 
loosing predictivity. Moreover, the connection among different decays is not straightfoward, a detailed discussion will be presented in Ref.~\cite{etaetap}. 
On the other hand, the high-energy behavior obtained from pQCD 
---which imposes a behavior on the TFF which renders the integral in Eq.~\eqref{eq:loopint} finite--- cannot be extrapolated down to such low energies. 
At the end, one is compelled then to use some phenomenological model accomodating (some of) the previous properties the best one can. 
This is the example of vector meson dominance (VMD) models~\cite{Bergstrom:1983ay}, quite used in the past, or the newer large-$N_c$ inspired ones~\cite{Knecht:1999gb,Husek:2015wta}. 
However, such approaches entail large model-dependencies~\cite{Masjuan:2015lca,Masjuan:2007ay}, which have been tried to ammend through the use of VMD 
data-fits~\cite{Dorokhov:2007bd}, at least for the single-virtual part, since no experimental data is availale for the doubly-virtual TFF.
The question remains then, how reliable these calculations are ---specially regarding their error evaluation--- and, given the current disagreement of the measured BR, to which 
extent such discrepancy is real.
Moreover, the approximation $\Lambda_{\textrm{TFF}}\gg m_P$ is not accurate anymore despite intuition. Actually, from Eq.~\eqref{eq:loopint}, 
the relevant TFF region is $-m_P^2 < Q^2 < \infty$, in contrast to Eq.~\eqref{eq:loopapp}, which requires an accurate description both in the spacelike and the Dalitz-decay 
$P\rightarrow\gamma\overline{\ell}\ell$ kinematics. This implies, specially for the $\eta'$, the appearence of aditional hadronic thresholds invalidating the common use of 
the unitary bound.
New techniques then are required to better describte the TFF accounting for all possible low- and high-energy constraints and reproducing the available experimental data. 
This highly non-trivial endeavour is explored in the next section.

\section{Our approach: Canterbury approximants}

In Ref.~\cite{Masjuan:2012wy}, it was proposed that the single-virtual TFF $F_{P\gamma^*\gamma}(Q^2,0)$ can be described through the use of Pad\'e approximants (PA)~\cite{Baker} 
in the space-like region, an idea that was extended to the $\eta$ and $\eta'$ cases in Ref.~\cite{Escribano:2013kba}. Pad\'e approximants, $P^N_M(Q^2)$ are rational 
functions of polynomials of degree $N$ and $M$, constructed in such a way that they match the series expansion of the function to be approximated up to 
$\mathcal{O}((Q^2)^{N+M+1})$, i.e., for the $F_{P\gamma^*\gamma}(Q^2)=F_{P\gamma^*\gamma}(0)(1+b_PQ^2+c_PQ^4+...)$, its PA reads
\begin{equation}
\label{eq:pa}
P^N_M(Q^2) = \frac{T_N(Q^2)}{R_M(Q^2)} = F_{P\gamma^*\gamma}(0)(1+b_PQ^2+c_PQ^4+...+\mathcal{O}(Q^2)^{N+M+1}).
\end{equation}
The advantage of PA is that they are able to extend the radius of convergence of conventional series expansions, for meromorphic, as well as for Stieltjes functions~\cite{Baker}, 
to the whole complex plane except at regions where the function is already ill-defined, such as poles or branch cuts. Therefore, it is expected both, from the point of view of the large-$N_c$ 
limit of QCD, and the analytic structure at low- and high-energies ---which is of the Stieltjes kind--- that PA converge, providing then a model-independent approach for describing the TFF 
once the low-energy parameters in Eq.~\eqref{eq:pa} are given. In Refs.~\cite{Masjuan:2012wy,Escribano:2013kba}, it was explained how to extract them in a 
model-independent way through the use of experimental space-like data with the inclusion, for the first time, of a systematic error. 
In addition, the use of two-point PA~\cite{Baker} allows to reconstruct the TFF from the low- and high-energy expansion at the same time. Actually, this property was used in 
Ref.~\cite{Escribano:2013kba} to determine the $\eta,\eta'$ asymptotic behaviors. 
With this description at hand, we observed that PA provide an excellent description for the $\eta$ Dalitz decay as well~\cite{Aguar-Bartolome:2013vpw} and very recent 
results from BES-III~\cite{Ablikim:2015wnx} show similar agreement for the $\eta'$. 
We used this time-like data for the $\eta$~\cite{Escribano:2015nra} and $\eta'$~\cite{etap} on top of the available space-like data in the first ever combined description 
in order to extract the most precise determination for the low-energy parameters to date.
In conclusion, our approach provides, for the first time, an excellent description for the TFF in the region of interest for this processes.\\

There is still the point, which goes beyond the present discussion, specially for the $\eta'$, on how to interpret the resonances appearing in the Dalitz region: 
these resonances produce an additional contribution to the imaginary part  
on top of the $\gamma\gamma$ one. From Pad\'e theory, we know that PA should effectively reproduce resonances and threshold-discontinuities contributions as well. 
These assessments are confirmed in our preliminary studies in Ref.~\cite{etaetap} and suggest that the unitary bound is lowered by $1\%$ and $40\%$ for the  $\eta$ and $\eta'$, respectively.
Further details are unimportant for the current discussion and will be given in Ref.~\cite{etaetap}.

The last point remaining then is the extrapolation from the single-virtual TFF to the doubly-virtual case with the current lack of data. This is achieved through the use of 
Canterbury approximants (CA), an extension of Pad\'e approximants to bivariate functions. They are defined analogous to PA, Eq.~\eqref{eq:pa}, as to match the original 
series expansion, 
\begin{align}
\label{eq:ca}
C^N_M(Q_1^2,Q_2^2) = \frac{T_N(Q_1^2,Q_2^2)}{R_M(Q_1^2,Q_2^2)} = F_{P\gamma^*\gamma^*}(0,0)(1 & +b_P(Q_1^2+Q_2^2) +c_P(Q_1^4+Q_2^4)+ \nonumber \\ 
 & +a_{1,1}Q_1^2Q_2^2 +...+\mathcal{O}(Q_1^2)^{\gamma}(Q_2^2)^{N+M+1-\gamma}),
\end{align}
with $\gamma \in (0,N+M+1)$, see~\cite{Masjuan:2015lca} and references therein. The simplest element reads~\cite{Masjuan:2015lca}
\begin{equation}
C^0_1(Q_1^2,Q_2^2) = \frac{F_{P\gamma^*\gamma^*}(0,0)}{1-b_P(Q_1^2+Q_2^2) + (2b_P^2-a_{1,1})Q_1^2Q_2^2}, 
\end{equation}
where $b_P$ is the slope from the single-virtual TFF series expansion, and the new parameter $a_{1,1}$ is the doubly-virtual slope.
Again, similar convergence theorems exist for the bivariate case which ensures the model-independency of the method~\cite{Masjuan:2015lca}. 
The problem we face in this case is that no data is available to extract those parameters as done in Refs.~\cite{Masjuan:2012wy,Escribano:2013kba,Escribano:2015nra},  
which is the reason for which we do not consider higher approximants, such as $C^1_2(Q_1^2,Q_2^2)$, which would allow to test the convergence of the $C^N_{N+1}(Q_1^2,Q_2^2)$ 
sequence. Consequently, we need an estimate for the $a_{1,1}$ parameter which may be thought as an educated guess  
generous enough as to cover its real value. To our best knowledge, there are two available insights. On the one hand, at very low-energies, there are indications suggesting 
that non-factorizable effects are suppressed~\cite{Bijnens:2012hf},
implying $F_{P\gamma^*\gamma^*}(Q_1^2,Q_2^2) \sim F_{P\gamma^*\gamma}(Q_1^2,0)\times F_{P\gamma^*\gamma}(Q_2^2,0)$, and $a_{1,1}=b_P^2$. On the 
other hand, the high-energy behavior states that, for large virtualitites, the TFF behaves as $\lim_{Q^2\to\infty} F_{P\gamma^*\gamma^*}(Q^2,Q^2) \sim Q^{-2}$, meaning that 
$a_{1,1}=2b_P^2$. This suggests that the real value should lie within $a_{1,1}\in\{b_P^2,2b_P^2\}$ range\footnote{A very recent analysis, Ref.~\cite{Xiao:2015uva}, seems to confirm 
the assumption we employed in Ref.~\cite{Masjuan:2015lca} and  here. Namely, that factorization should hold at low-energies for the $\eta$ TFF at least. Note however that we 
do not assume this holds true at a larger energy range as this study suggests.}. 
Of course, experimental data or lattice QCD will have the last word on this. For the moment, we take this range for our calculations as the best estimate to be on 
the conservative side. Notice here that such information plays a marginal role at low energies where 
$\tilde{F}_{P\gamma^*\gamma^*}(Q_2^2,Q^2) \simeq C^0_1(Q^2,Q^2) =  1 -2b_PQ^2 + (2b_P^2+a_{1,1})Q^4 + ... $ as the $Q^2$-term dominates there. To illustrate this, using this 
expansion and integrating in Eq.~\eqref{eq:loopapp} up to a cut-off $\Lambda$, we find that factorization or OPE results are barely the same up to $\Lambda=0.4$~GeV, with 
$\textrm{BR}(\pi^0\rightarrow e^+e^-)=6.64\times10^{-8}$. It will be necessary nevertheless to find whether OPE or factorization 
represents the better choice above if precision is required, particularly in the $\Lambda=(0.4-2)$~GeV region, with special emphasis on its lower-energy part. 
On the other hand, reproducing or not the experimental value will depend on an unexpected $a_{1,1}$ value which would be already observable at very low 
energies~\cite{Masjuan:2015lca}. Its eventual value could be then determined from this hypothetical data through the use of CA in a model-independent fashion.

\section{Results for $\textrm{BR}(P\rightarrow\overline{\ell}\ell)$}

In the following we present our numerical results. Besides implementing the correct low-energy behavior, we provide an exact calculation of Eq.~\eqref{eq:loopint} rather 
than~\eqref{eq:loopapp}, which respresents 
---specially for the $\eta,\eta'$ and $\mu^+\mu^-$ decays--- an additional improvement with respect to previous studies.
We show in Tab.~\ref{tab:res} our preliminary results from our chosen range $a_{1,1}=(2b_P^2\div b_P^2)$, see Ref~\cite{etaetap}\footnote{The 
above results are still preliminary and represent the work in progress from Ref.~\cite{etaetap}. Notice in addition the different range to that chosen in Ref.~\cite{Masjuan:2015lca}. 
The reason for our previous choice was meant to illustrate implications from experimental $\pi^0\rightarrow e^+e^-$ results which is of no relevance here.}.
\begin{table}
\centering
  \begin{tabular}{cccc} \hline
    Process & BR$^{\textrm{th}}$ & BR$^{\textrm{exp}}$ & Ref. \\ \hline
    $\pi^0\rightarrow e^+e^-$ & $(6.20\div6.35)(5)\times10^{-8}$ & $7.48(38)\times10^{-8}$ & \cite{Abouzaid:2006kk}\\ \hline
    $\eta\rightarrow e^+e^-$ & $(5.31\div5.44)(^{+4}_{-5})\times10^{-9}$ & $\leq2.3\times10^{-6}$ & \cite{Agakishiev:2013fwl} \\
    $\eta\rightarrow \mu^+\mu^-$ & $(4.72\div4.52)(^{+4}_{-8})\times10^{-6}$ & $5.8(8)\times10^{-6}$ & \cite{Abegg:1994wx} \\ \hline
    $\eta'\rightarrow e^+e^-$ & $(1.82\div1.86)(19)\times10^{-10}$ & $\leq5.6\times10^{-9}$ & \cite{Achasov:2015mek,Akhmetshin:2014hxv} \\
    $\eta'\rightarrow \mu^+\mu^-$ & $(1.36\div1.49)(33)\times10^{-7}$ & $-$ \\ \hline
  \end{tabular}
\caption{Preliminary results for the different decays in comparison to experimental results using Eq.~\protect\eqref{eq:loopint}. }
\label{tab:res}
\end{table}
Note that the high-energy condition decreases all the BRs with the 
exception of the $\eta\rightarrow\mu^+\mu^-$ decay. The errors appearing arise from the low-energy parameters 
determination as well as for the systematic error of the procedure, which is non-negligible as we cannot go beyond the simplest $C^0_1(Q_1^2,Q_2^2)$ approximant. 
For a detailed discussion on systematic errors, see Ref.~\cite{etaetap}. Notice that our preliminary estimates of our systematic error for the $\eta'$ seems to be rather large 
$(25\%)$, beyond the impact of the $a_{1,1}$ range. This very large error comes from the way we are accomodating resonance effects in our framework. Further work is undergoing and we expect to diminish it in Ref.~\cite{etaetap}.

Our method provides, for the first time, a serious tool to assess a theoretical error from the mathematical point of view, which seems to be higher than previously 
estimated~\cite{Knecht:1999gb,Dorokhov:2007bd,Dorokhov:2009xs,Bergstrom:1983ay}\footnote{We note that, when comparing to existing results ---even if all mass-corrections are taken--- 
discrepancies may appear if the values implied by their TFF slopes are different.}.
In addition, we find non-negligible 
corrections for the $\eta$ and $\eta'$ decays. Though this was noted before~\cite{Dorokhov:2009xs}, this is the first calculation providing 
the exact result. Moreover, our approach is the only one able to make full use of experimental data on both, the space-like and low-energy time-like regions.

From Tab.~\ref{tab:res}, we find a $3\sigma$ discrepancy for the $\pi^0$. A recent reanalysis of radiative corrections~\cite{Vasko:2011pi,Husek:2014tna} suggests though a 
smaller BR of $6.87(36)\times10^{-8}$, reducing the discrepancy if their results are confirmed. Similarly, we find $1.3\sigma$ deviation for the $\eta\rightarrow\mu^+\mu^-$ decay.
Even if the discrepancies are not that large yet, the existing difference would imply either new physics, or values for the $a_{1,1}$ parameter out of 
theoretical expectations, i.e., an unexpected behavior of the doubly virtual TFF at low-energies.
This makes very interesting both, a new measurement of these decays with higher precision to discard a statistical fluctuation, and a first measurement of the 
doubly-virtual TFF, which would determine once for all the nature of the discrepancy. Moreover, given our obtained value for the 
$\eta'\rightarrow e^+e^-$ result and present bounds~\cite{Achasov:2015mek,Akhmetshin:2014hxv}, we would like to encourage our experimental colleagues in 
Novosibirsk to push further their measurement at SND and CMD-III to reach this limit, higher than naive expectations from the (wrong) unitary bound.

We stress that an unexpected behavior of these TFFs would have a large impact in the hadronic light-by-light piece of $(g-2)_{\mu}$, as it enters in the 
$\pi^0$ contribution~\cite{Masjuan:2015lca,Sanchez-Puertas:2015yxa}. We believe therefore that clarifying the current situation in $P\rightarrow\overline{\ell}\ell$ both on the 
theoretical and experimental side, represents an important task for any approach calculating the hadronic light-by-light contribution to $(g-2)_{\mu}$.

\section{Summary and outlook}

In this work we have revised the $P\rightarrow\overline{\ell}\ell$ decays for $P=\pi^0,\eta,\eta'$ in a model-independent way with the machinery of Pad\'e 
theory extended to the doubly-virtual case, Canterbury approximants.
Such approach is data driven and allows, for the first time, to include the space-like and low-energy time-like data together with the low- and high-energy constraints 
as well as a systematic errors.  
Moreover, our approach provides a tool to systematically analyze experimental data on the doubly virtual transition form factors beyond the standard modelization procedures 
used at experiments.
Given the existing discrepancies in all the measured channels, we have studied the role of double virtuality with emphasis in the low-energy domain 
---which has been overlooked--- without negelcting the high-energy behavior. 
In addition, we have performed the exact numerical calculation for these decays, which is specially relevant for the $\eta$ and $\eta'$.
We have found that our results cannot acommodate the experimental results unless the low-energy behavior of the TFF is out of expectations. To confirm this, data 
on the doubly virtual TFF would be required. In addition, new precise measurements for these processes would be very welcomed in order to discard 
any statistical fluctuation and explore the presence of new physics. Moreover, these implications are of interest for the community involved in $(g-2)$ physics.

\section{Acknowledgements}

We would like to thank Marc Vanderhaeghen for encouragement and discussions. Work supported by the Deutsche Forschungsgemeinschaft DFG through 
the Collaborative Research Center ``The Low-Energy Frontier of the Standard Model'' (SFB 1044).

\end{document}